\documentclass{vgtc}                          

\ifpdf
  \pdfoutput=1\relax                   
  \usepackage{graphicx}                
  \DeclareGraphicsExtensions{.pdf,.png,.jpg,.jpeg} 
\else
  \usepackage{graphicx}                
  \DeclareGraphicsExtensions{.eps}     
\fi%

\graphicspath{{figures/}{pictures/}{images/}{./}} 

\usepackage{microtype}                 
\PassOptionsToPackage{warn}{textcomp}  
\usepackage{textcomp}                  
\usepackage{mathptmx}                  
\usepackage{times}                     
\usepackage{cite}                      
\usepackage{tabu}                      
\usepackage{booktabs}                  

\onlineid{2137}

\vgtccategory{applications}

\vgtcinsertpkg

\usepackage[dvipsnames]{xcolor}
\usepackage{soul}
\usepackage{subcaption}
\usepackage{comment}
\usepackage[utf8]{inputenc}

\usepackage[]{algorithm2e}
\usepackage{amsmath}
\usepackage{todonotes}

\usepackage{kotex}
\usepackage{color}
\definecolor{blue}{rgb}{0,0,0.6}
\definecolor{green}{rgb}{0,0.3,0}
\definecolor{red}{rgb}{0.6,0,0}
\definecolor{gray}{rgb}{0.4,0.4,0.4}
\definecolor{purple}{rgb}{1,0,1}

\usepackage{enumitem} 

\title{A Full Body Avatar-Based Telepresence System for Dissimilar Spaces}

\author{Leonard Yoon\thanks{e-mail: lyoon@kaist.ac.kr}\\ %
        \scriptsize Korea Advanced Institute of Science and Technology %
\and Dongseok Yang\thanks{e-mail: dsyang@kaist.ac.kr}\\ %
     \scriptsize Korea Advanced Institute of Science and Technology %
\and Choongho Chung\thanks{e-mail: thegenuine@kaist.ac.kr}\\ %
     \scriptsize Korea Advanced Institute of Science and Technology %
\and Sung-Hee Lee\thanks{e-mail: sunghee.lee@kaist.ac.kr}\\ %
     \scriptsize Korea Advanced Institute of Science and Technology}
\teaser{
   \centering
   \includegraphics[width=1.0\columnwidth]{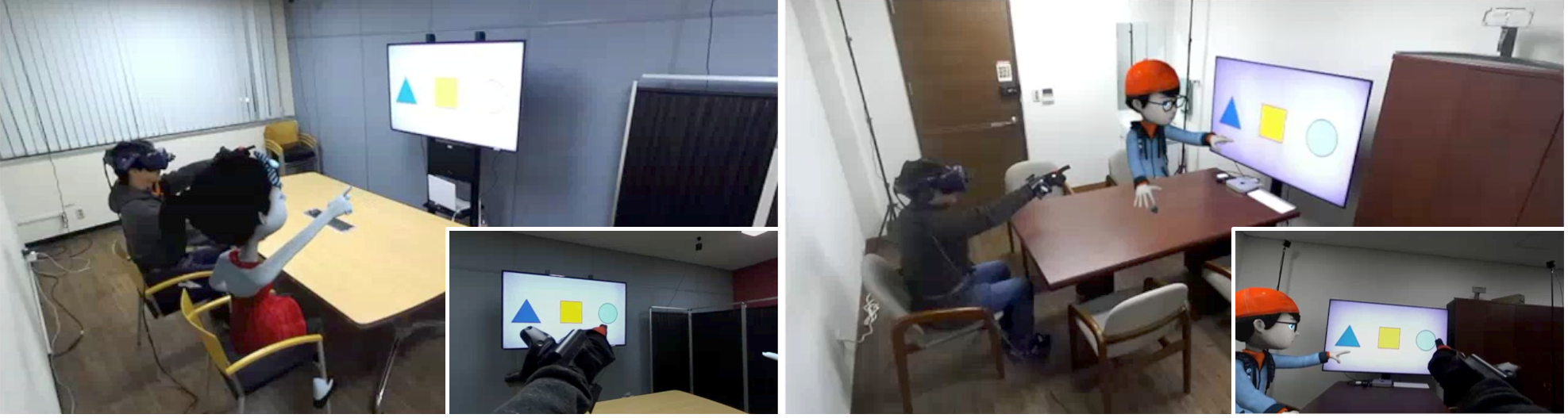}
   \caption{Our telepresence system allows for different room environments by retargeting the user's placement and deictic gesture to the remote avatar so as to preserve the user's context of placement and interaction. In this figure, a user in the left space is sitting facing a display and pointing at a point in the display. The user is represented in the right space by a male avatar  sitting on the right side of a display, pointing at the corresponding point in the display.  A user sitting in front of the display in the right space is represented by a female avatar sitting to the left of the display in the left space.
   Small figures are user's egocentric views.}~\label{fig:teaser}
}

\abstract{We present a novel mixed reality (MR) telepresence system enabling a local user to interact with a remote user through full-body avatars in their own rooms. If the remote rooms have different sizes and furniture arrangements, directly applying a user's motion to an avatar leads to a mismatch of placement and deictic gesture. To overcome this problem, we retarget the placement, arm gesture, and head movement of a local user to an avatar in a remote room to preserve a local user's environment and interaction context. This allows avatars to utilize real furniture and interact with a local user and shared objects as if they were in the same room. This paper describes our system's design and implementation in detail and a set of example scenarios in the living room and office room. A qualitative user study delves into a user experience, challenges, and possible extensions of the proposed system.} 

\CCScatlist{
  \CCScatTwelve{Human-centered computing}{Collaborative and social computing}{Collaborative and social computing systems and tools};
}

\begin{document}


\firstsection{Introduction}

\maketitle

Recent advances in technology facilitated room-scale telepresence research in mixed reality (MR) setting, which allows a user to communicate with a remote partner augmented as an avatar. If two users want to communicate with each other, augmenting a virtual copy of a remote user and some nearby objects at the same location in the local space can deliver the context of the user's motion \cite{orts2016holoportation}.

However, in the case of using a whole space of two rooms with different shapes and object arrangements, directly placing an avatar with respect to the spatial relation between a user, an avatar, and shared objects of interest may cause a discrepancy, for instance, an avatar sitting in the air or penetrating some objects. To avoid these artifacts, the avatar may be only allowed to be placed on limited areas such as a free space \cite{Lehment2014} or a sofa \cite{Pejsa:2016}. The drawback is that an available communication space is restricted to only a sub-space of the whole space. Alternatively, using a partial representation (e.g., upper body) of an avatar can reduce the mismatch of environmental context between different spaces \cite{spatial}. This is effective for the typical tasks in an office meeting, but the absence of full-body motion limits the scope of interaction and reduces the presence of the partner.

To overcome these limitations, we can compromise the convenience of the sameness of spatial arrangement of a user, an avatar, and objects across spaces. Instead, place a full-body avatar to best match a user's surrounding environment for interaction with a partner and shared objects. To this end, we need to solve two major issues. First, find an optimal placement of an avatar to match the environmental context and interaction context as much as possible. Second, generate avatar motions to keep the interaction context and deal with spatial relation between a user, an avatar, and a shared object according to space.

The avatar placement from the different environment of two spaces has been studied by \cite{Jo:2015, Placement}. However, to our knowledge, a telepresence system that places an avatar to an optimal location and animates it to keep the interaction context has not been developed yet. This paper presents our telepresence system to achieve this goal. In the system, we place the avatars by using the method of \cite{Placement}. Besides, we newly develop an avatar gesture retargeting method that detects a deictic pointing gesture of the user and retargets it to the avatar. Combining the two enables the users to move between different locations of a small room to sit and stand while interacting with a partner using shared objects. 

Figure \ref{fig:teaser} shows a snapshot of using our telepresence system. Confronting the difference in the rooms' furniture arrangement, the avatars take the appropriate locations that accommodate sitting pose and make pointing gestures towards the same target as the users point at. The proposed system has been implemented with off-the-shelf consumer products. After presenting the system's details, we show use case scenarios in the office room and living room. Lastly, we conduct an exploratory user study to learn about the user experience and possible extension direction.

\section{Related Work}
This section introduces previous work on avatar-based telepresence systems and on creating deictic gestures in a telepresence environment.

\subsection{Mixed Reality Telepresence Systems}
Inspired by pioneering work for telepresence system \cite{fuchs1994virtual, kurillo2008immersive, petit2010multicamera, beck2013immersive}, many researchers have continuously challenged to improve the sense of co-presence with the development of HMD-based MR device. A group of seminal work focused on real-time 3D spatial capture of a person's surrounding scene in remote space, which was augmented in a local space \cite{maimone2013general, orts2016holoportation}. Another direction of research has found values in investigating a shareable space between two remote rooms for conferencing \cite{Lehment2014}, which was later extended to generating workspace for multi-room with a complex furniture arrangement \cite{keshavarzi2019optimization}. On the other hand, our system aims to use a whole space of two rooms by placing an avatar in each space for people naturally residing in different spaces.

There have also been studies on the user's placement in a certain remote space location to enhance communication and collaboration \cite{Jo:2015, Pejsa:2016}. However, these systems were limited to maintaining an environmental context, mainly a sitting affordance, with a body pose adjustment or a virtual mirroring of the projected person to match the interaction context. On the other hand, our system adjusts the full-body motion of avatars in each space to maximize the environmental and interaction context. Recently, Piumsomboon et al. \cite{piumsomboon2018mini} presented a reduced sized avatar with gaze and body gesture for remote collaboration, and Kumaravel et al. \cite{thoravi2019loki} proposed a mixed reality system for teaching physical tasks using a partial representation avatar with 3D spatial capture as well as audio and 2D video. While these studies focused on collaborating within a shared sub-space or teaching that occur in only one space at a time, we utilize both spaces simultaneously for a broader range of tasks as discussed in Sec. \ref{scenario}.

\subsection{Deictic Pointing in Telepresence}
Many previous studies on telepresence incorporated deictic pointing in various settings. Two contrasting approaches can be identified for configuring telepresence applications that enable pointing interactions, emphasizing using either virtual or physical avatars for representing remote users. 

The first approach makes use of virtual avatars for pointing activities. Beck et al. \cite{beck2013immersive} demonstrated a group-to-group virtual collaboration system that projects remote user avatars by aligning both systems' coordinates to match avatar locations, aided in reducing pointing errors. In another work by Higuchi et al. \cite{higuchi2015immerseboard}, pixelated touch screens were used as shared referential input devices. The screen doubled as a display for a 3D reconstructed user avatar whose arm can be stretched to provide pointing metaphors to the other user. More recent approaches also made use of augmented space for collaborations and virtual avatars for pointing interactions. Such examples include the use of scaled-down avatars for collaborations in MR environment for AR headset users \cite{piumsomboon2018mini}, or generation of deictic gesture animations for avatars from speech and text input for highlighting key phrases projected on a virtual screen \cite{kappagantula2019automatic}.

The second approach explores deictic gestures for remote environments when using teleoperated robots as user avatars. For humanoid robots, deictic pointings were demonstrated as an interaction mode between two identical or symmetric physical spaces \cite{nagendran2015symmetric}. For non-humanoid robots in MR settings, exploiting the natural human context of pointing is more difficult due to the different shapes of human operators and robots. Hence, more flexible interpretations of deixis or pointing behaviors in various perspectives were suggested \cite{williams2018framework, williams2019mixed} to highlight objects of importance in the user's space. Although not directly tackling the natural pointing motion retargeting between different spaces, it is evident that the physical manifestation of a user avatar results in more active participation in social interactions \cite{stahl2018social, kim2017large}.

There is an increasing focus on MR collaborations with virtual workspace and avatars for interactions regarding the use of deixis between remote users. We aim to showcase uses of deictic pointings in MR scenes further, where communication is enabled between two dissimilar spaces in real-time, and avatars' deictic gestures are adaptively retargeted from hand tracking information of users to enable more accurate and natural exchanges of non-verbal cues.

\begin{figure}[t]
\centering
  \includegraphics[width=1.0\columnwidth]{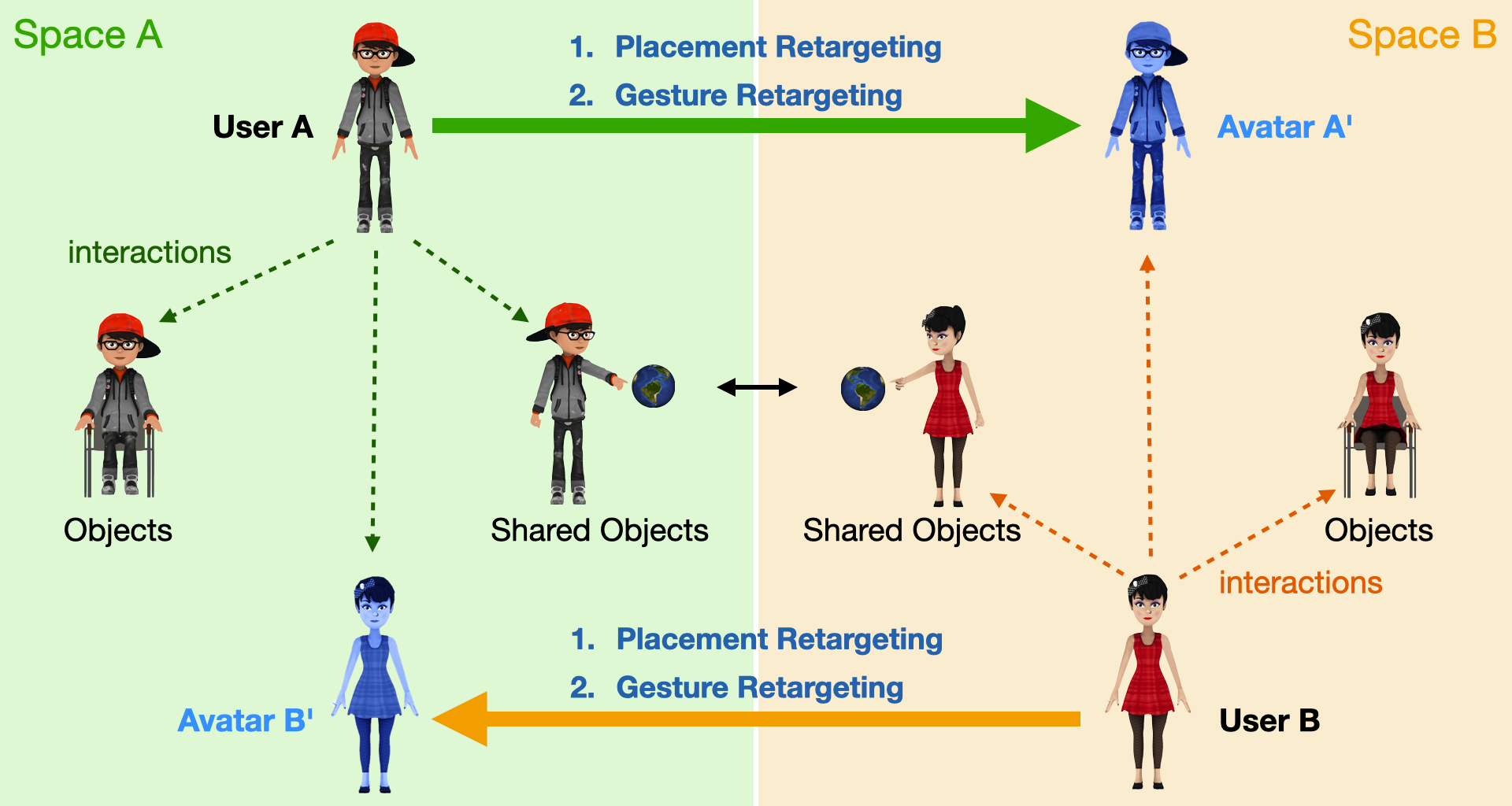}
  \caption{Telepresence system for dissimilar spaces. Augmented user avatars undergo placement and gesture retargeting for communications between two dissimilar spaces A and B.}~\label{fig:design}
\end{figure}
\begin{figure*}[ht]
\centering
  \includegraphics[width=2.0\columnwidth]{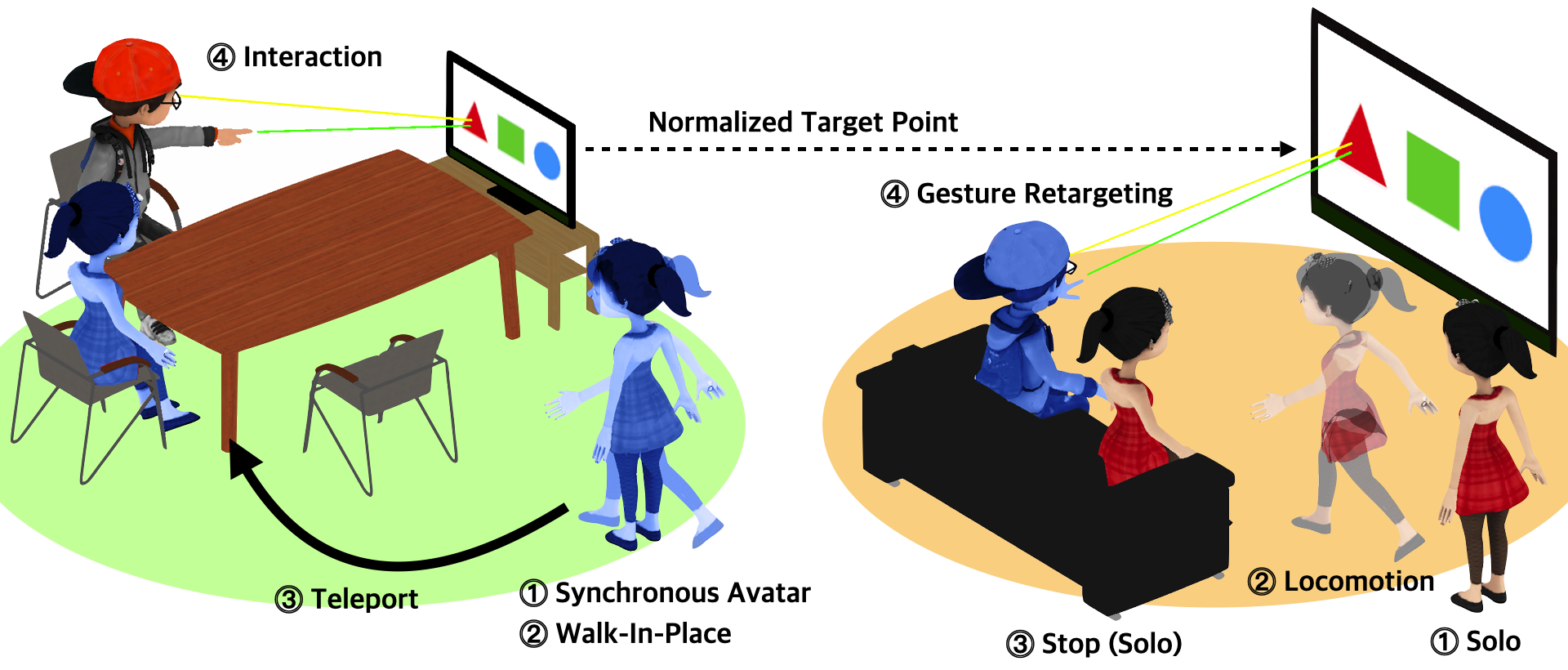}
  \caption{Overview of avatar (blue) motion generation. Location change of the user through walking (right, \textcircled{1} $\rightarrow$ \textcircled{3}) is retargeted to walk-in-place and teleportation of avatar. Deictic gesture of the user (left, \textcircled{4}) is retargeted to preserve the interaction context in the remote space. }~\label{fig:overview}
\end{figure*}
\section{Designing Telepresence FOR Dissimilar spaces}
Figure \ref{fig:design} illustrates the concept of user actions and avatar animation in our telepresence system. Our system's goal is to enable a user to make use of his or her own space, including existing furniture, while interacting with a remote person represented as a full-body avatar while sharing a screen or augmented virtual objects. 
To deliver the user's environmental and interaction context in the local space to the remote space, the user's motion needs to be retargeted to his or her avatar. With this goal in mind, we designed our system as follows:
\begin{itemize}
     \setlength\itemsep{0.5pt}
     \item The avatar is placed to match the environment context of the remote user and the spatial relation with partner.
    \item The avatar's upper body gesture is modified to preserve the interaction context of the remote user if the user is paying attention to interaction targets. 
     \item Candidate interaction targets include the partner's head, shared screen, and augmented virtual objects for pointing and gazing activities.
\end{itemize}

The main challenge is to find an optimal placement of an avatar that preserves the multiple environmental contexts at the same time. This is challenging due to the different sizes and configurations of the two spaces. For example, if a user in a local space is sitting in a chair in front of a partner while watching a TV, there might not be a perfect location in remote space to match all three contexts because simply placing an avatar with respect to one of the three contexts will hinder seamless communication and collaboration. To this end, we exploit an avatar placement algorithm that takes into account environmental and interaction contexts as much as possible.

The second challenge is to match the deictic context of pointing gestures. Because the spatial arrangements of partners and shared objects are different by space, directly synchronizing user motion to the avatar can relay inaccurate information to the other user in the remote space. To prevent this, a user's deictic gesture should be detected in the user's local space and then adjusted to the remote space for the augmented avatar while preserving its deictic context.

By integrating placement and deictic gesture retargeting for avatars in dissimilar spaces, we can reduce confusion and ambiguity in interpreting both spatial and deictic context of the other user who is communicating from a different physical space through a virtual avatar. A detailed explanation of avatar placement and gesture retargeting is described next.

\section{System}
Our system realizes telepresence between two rooms using a full-body avatar that either mimics the (synchronous) motion of a real person or retargets the user's motion based on the user's predefined states. The user's state is divided into \textit{locomotion}, \textit{interaction}, and \textit{solo} (see Figure \ref{fig:overview}). The locomotion state is when the user is moving from one point to another. If the user's gaze or hand direction is fixated on one of the interaction targets, the user is in the interaction state. Otherwise, the user is in a solo state. We describe object models of our telepresence system first and then explain each user state's recognition and the corresponding strategy for avatar movement.

\subsection{Objects in Space}
Our placement and retargeting algorithms require coordinates and labels of objects (e.g., sittable objects, screens, and pointing target) in the real scene. As our system does not consider real-time 3D reconstruction and segmentation, we used representative 3D virtual object models (e.g., chair, sofa, table, and TV) and manually placed, scaled, and labeled them to match the real objects. From this invisible virtual representation of real spaces, we obtain raw data to form a feature vector for avatar placement and target pointing point for gesture retargeting.

Among the objects in the spaces, only those that form a matching pair for both spaces are considered candidate objects of interaction, including tables, TVs, and a user-avatar pair. Objects in the pair are not necessarily of the same size; when an object is pointed, we obtain a normalized local coordinate of the object with respect to the object's size. We then scale the normalized coordinate with respect to the paired object's size in the remote space to obtain the target pointing location for the avatar.

\subsection{Synchronous Avatar}

If the user is in the solo state, his or her pose is directly applied to the avatar. For this, the position and orientation of the hands, feet, and head relative to the root are measured from the user and transmitted to the avatar as its goals. Then, an inverse kinematics (IK) solver computes the avatar's whole body pose to achieve the goals. In addition to that, finger joint rotations are captured from finger tracking gloves and directly applied to the character model's finger joints. This allows the user to utilize full-body gesture interaction, including deictic gestures for non-verbal communication.

\subsection{Placement of Avatar}
\begin{figure}
\centering
  \includegraphics[width=1.0\columnwidth]{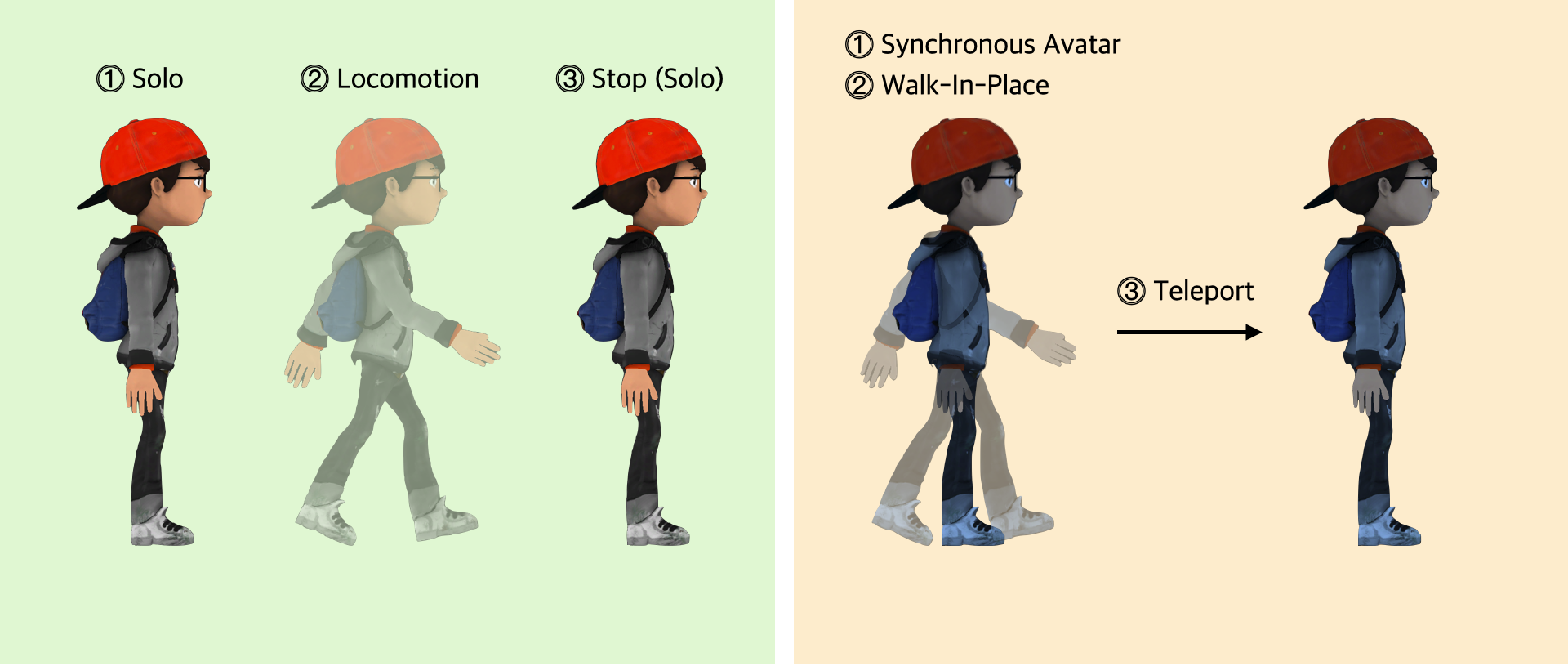}
  \caption{Avatar placement strategy representing user locomotion}~\label{fig:placement}
\end{figure}

When the user starts to move with a speed (of the pelvis) over a fixed time frame ($\approx 166$ ms) exceeding a locomotion threshold, the state of the avatar changes from solo to locomotion. During locomotion, the avatar's motion turns to Walk-In-Place (WIP); the avatar's position and orientation do not change, but the avatar's pose mimics that of the user, resulting in a still, walking motion. When the user's pelvis speed in locomotion state reaches below the stop threshold, the avatar is teleported to a new placement (see Figure \ref{fig:placement}). The benefit of the WIP strategy is that it preserves the context of moving while hiding the user's actual moving path, which is not meaningful and sometimes seems unrealistic in the remote space because of unnatural penetrations of remote user and furniture.

Due to the different sizes and configurations of two spaces, an avatar's placement in remote space and a user's location in real space cannot be mapped one-to-one. Alternatively, we use an avatar placement algorithm from \cite{Placement}, which finds the avatar's placement in a dissimilar space as \textit{similar} as possible to the user's placement. Here we briefly summarize the algorithm.

The similarity between two placements is represented with several features as follows:
\begin{itemize}[noitemsep]
\item[--]  \textbf{Interpersonal relation:} Relative position and orientation between partners.
\item[--]  \textbf{Pose accommodation:} Height map of a user's intimate space (0.5m radius), representing the feasibility of accommodating user taken poses, such as sitting or standing.
\item[--]  \textbf{Visual attention:} category and distance of objects within a narrow (40$^{\circ}$) FOV of user. 
\item[--]  \textbf{Spatial feature:} category and distance of objects within a user's social space (3m radius).
\end{itemize}
In general, no placement can perfectly satisfy all the features' similarities, so we choose a placement that best compromises among different importance of features. 
Because the importance of individual features varies depending on space and user context, a deep neural network is trained to output the single integrated similarity value between two placements.
A triplet loss framework is used to learn the similarity 
from user preference data of avatar placement on different configurations of the room and location of a partner.

Given a user placement, our telepresence system samples a room as a coarse 2D grid (with the size of 0.25 meter and 24 orientations) followed by a finer sampling using the particle swarm optimization (PSO) \cite{eberhart2000comparing} to find an optimal placement that gives the highest similarity value. As a result, our system places the avatar to preserve the environment context (e.g., sit on a chair or stand in front of a screen) and/or interpersonal relations (e.g., to face each other or to watch a TV together) as much as possible.

After the avatar is teleported to the optimal placement in the remote space, the user state returns to the solo. 

\subsection{Interaction with Partner and Shared Objects}

\begin{figure}
\centering
  \includegraphics[width=1.0\columnwidth]{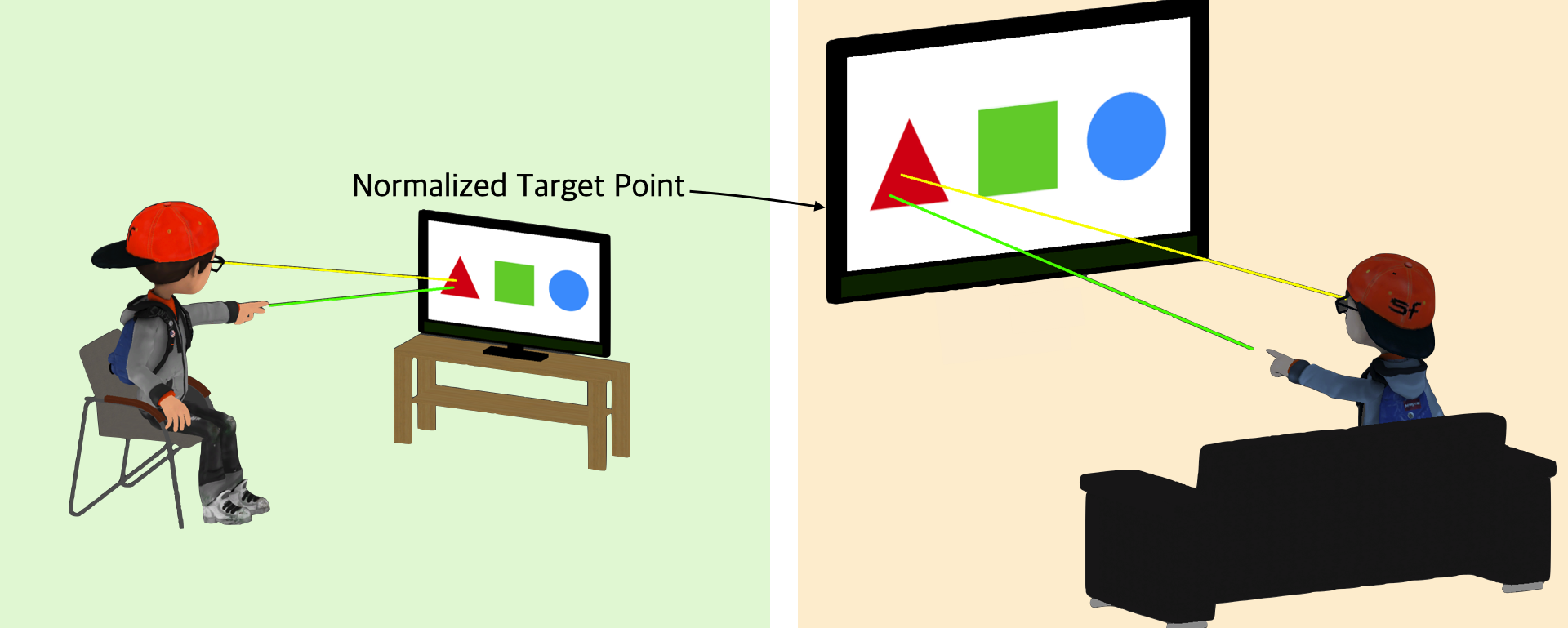}
  \caption{Deictic gesture retargeting enables a user in local space (left), and user avatar in remote space (right) to point at the same corresponding location of a shared object.}~\label{fig:gesture}
\end{figure}

If the user engages in interaction with the partner or the shared virtual object, the state changes to the interaction, and the avatar's movement is decoupled from the user's and made to realize the interaction context in the remote space as shown in Figure \ref{fig:gesture}. 
To this end, our telepresence system recognizes the user's interaction state and provides a suitable retargeting strategy for the avatar.

\subsubsection{Detecting Interaction and Interaction Target}

Our system determines that the user is interacting with other entities when his or her gaze or hand pointing directions are fixated to one of the candidate interaction targets. To this end, the system first detects collisions by raycasting from an HMD and trackable VR gloves. For the hand, a collision ray is cast forward from the tracked VR hand model to detect pointed objects from the hand while a center-eye raycasting from HMD detects gazing from the head. If a collision with a candidate is continued over a certain fixation threshold time, the user state is changed to the interaction state and the candidate entity is registered as an interaction target for the affected end effector.

\begin{figure}
\centering
  \includegraphics[width=1.0\columnwidth]{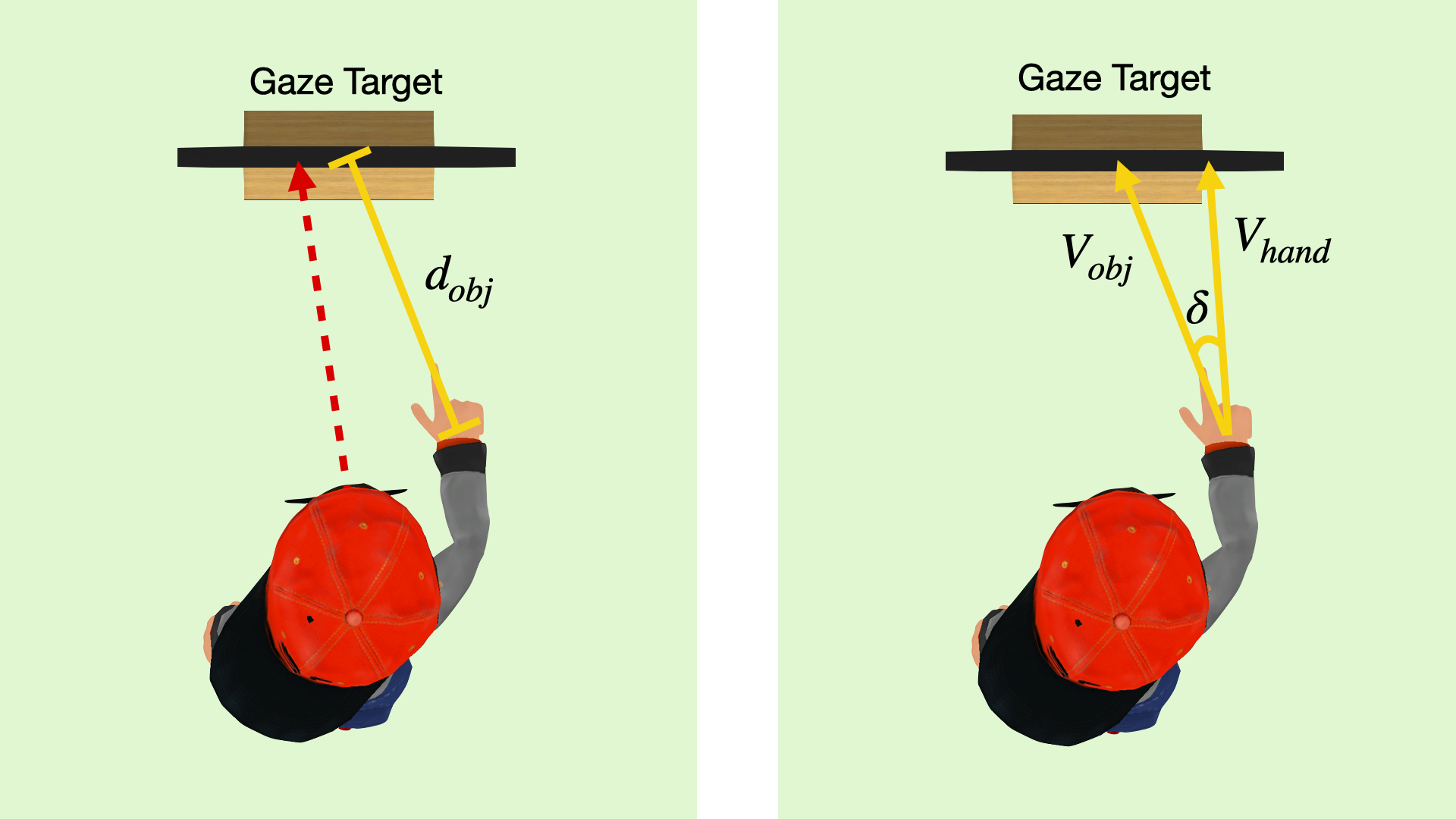}
  \caption{Conditions for matching hand interaction target to gaze target. When a hand moves toward the gaze target constantly over a threshold velocity (left), and the hand vector converges constantly to the object vector over threshold angular velocity (right) for certain period, we assume that the hand is moving to point to the gaze target.}~\label{fig:conditions}
\end{figure}

In addition, to decrease latency in fixation detection, we consider common user tendencies to gaze at a target first before pointing to it when performing a deictic gesture. We start by checking if any fixated target exists for the head. In such a case, we then observe an average decrease of distance ${\overline{\Delta d}}_{\text{obj}}$ between the hand to the gaze target, and an average rate of angle $\overline{\Delta \delta}$ decrease between hand forward vector and a vector from the hand to the gaze target  (Figure \ref{fig:conditions}).
\begin{align}
    {\overline{\Delta d}}_{\text{obj}} < V_\text{ threshold} < 0 &&\text{(Distance Condition)}
\end{align}
\begin{align}
    \overline{\Delta \delta} < {\omega}_\text{ threshold} < 0 &&\text{(Angle Condition)}
\end{align}
If both values remain smaller than respective threshold values for a fixed period of time, we assign gaze target as the hand's target. Algorithm \ref{alg:target acquisition} provides the procedure to identify target objects of the head and hands.

The local position of collision with respect to the target object's frame of reference is relayed to the other corresponding object in the other space. Normalized local coordinates of pointing are transferred to the remote space and rescaled for the counterpart object to provide the correct pointing location for the avatar. 

\begin{algorithm}
\KwResult{head\_targetobj, hand\_targetobj}
hand\_targetobj = null\;
head\_targetobj = raycast(head)\;
\For {for left hand, right hand}{
    \If{hand is lifted}{
        hand\_targetobj = raycast(hand)\;
    }
    \If{head\_targetobj != null}{
        \If{head\_targetobj != hand\_targetobj}{
            \If{${\overline{\Delta d}}_{\text{obj}} < V_\text{ threshold}$ and ${\overline{\Delta \delta}} < {\omega}_\text{ threshold}$}{
            hand\_targetobj = head\_targetobj\;
            }
        }
    }
}
\caption{Overview of user interaction target acquisition. Head and hands are allowed to have different interaction targets.}
\label{alg:target acquisition}
\end{algorithm}

\subsubsection{Gesture Retargeting}
Given the acquired gaze and hand pointing targets and the corresponding set of target coordinates, we control the avatar's head and arm to realize the targets by setting their desired IK goals. The IK goal of the head orientation is set to look at the target point. For the arm, the IK goal is set for its forward vector (shoulder to wrist joint) to point at the target while preserving the user's elbow flexion. 
The hand is rotated to point to the target while its up vector is set equal to that of the user's hand to retain its posing style. 

While an exact match in pointing vector towards the corresponding target like in Figure \ref{fig:gesture} seems ideal at first, it was found that humans have tendencies to perceive pointing location lower than the actual site of pointing \cite{wong2010you}. We compensate for this drop in pointing location by adjusting the arm pose, considering observations from previous works \cite{herbort2016spatial, sousa2019warping}. Since our main scenarios considered face-to-face or side-by-side interactions while assuming close proximities (less than 2m) between the user and pointing targets, we did not apply any horizontal post adjustments for retargeted pointing gestures in our work.

When the user state changes between solo and interaction, abrupt changes of the head and arm pose a negative impact on the naturalness of the avatar motion, which is overcome by smoothly interpolating between the current and desired poses. For this, we employ spherical linear interpolation for the head orientation for its forward vector to point at the target. We use a Bezier curve for smooth interpolation of the avatar hand defined by the avatar hand position and its forward vector as a tangent. The head and hands' target poses are updated in real-time as the user's target location undergoes small changes due to small movements of the head and arm. The interpolation speed is manually adjusted to fit our scenario that either aim for smoother transition or quicker arrival at the desired pose.

\subsection{Synchronous Audio}
Our system includes a real-time audio connection between the users in distant spaces. Two users can communicate verbally through the channel. The system does not support spatial sound for now.

\section{Implementation}
\begin{figure}
\centering
  \includegraphics[width=1.0\columnwidth]{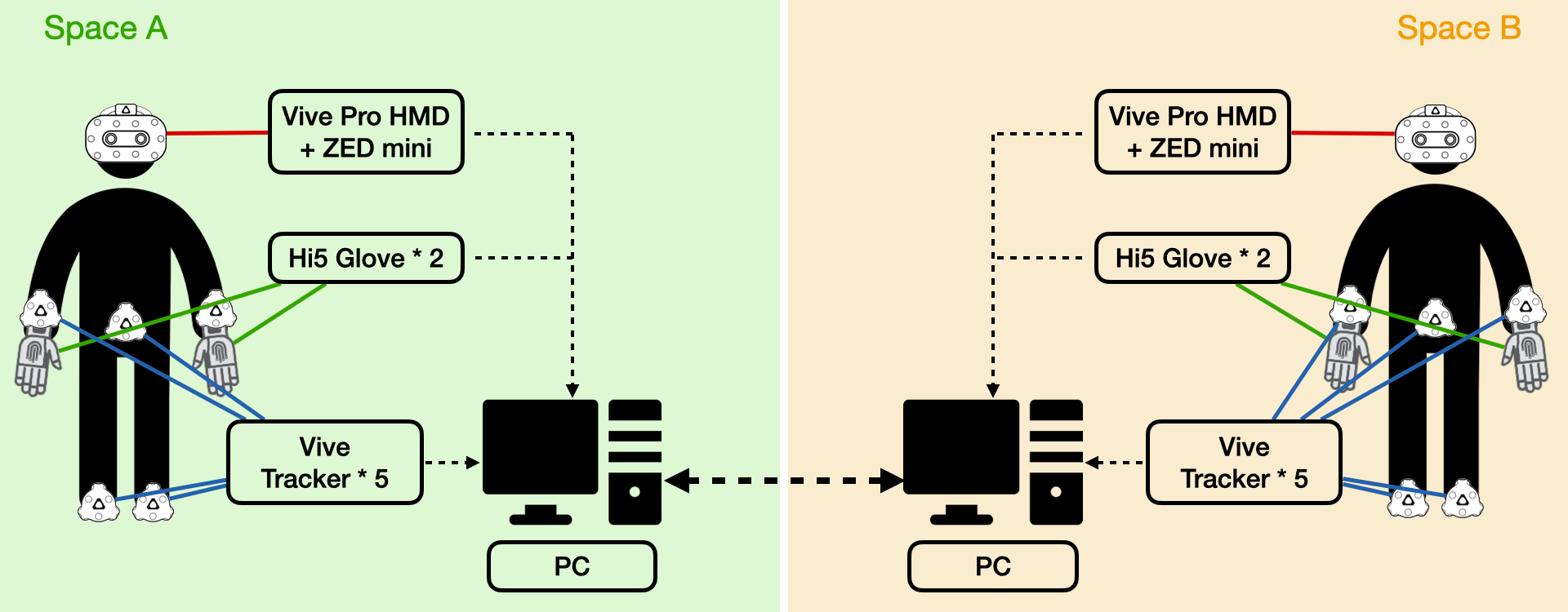}
  \caption{Overview of hardware configuration for interactions between two dissimilar spaces}~\label{fig:hardware}
\end{figure}

Our system consists of two identical hardware setups for each distant space (see Figure \ref{fig:hardware}). We integrated an RGB-D camera (ZED mini) on the HTC Vive Pro headset for MR rendering and real-time occlusion between real and virtual objects. A user wears additional Vive trackers to track joints and reconstruct the whole body pose with an IK solver (Final IK). The system is implemented with Unity3D engine (2019 2.6f1) and SteamVR framework. The placement module is implemented with TensorFlow and communicates with the system via Python socket connection. The whole system is running on PCs with Intel i9 processors and Nvidia Titan Xp graphics cards.

{\bf Mixed Reality Rendering.}
We used ZED mini stereo camera for mixed reality rendering for its resolution (up to 2.2k, 720p for our system) and field of view (90 degrees (H) x 60 degrees (V)). The ZED mini's depth-sensing ranges from 0.1m to 15m, enabling real-time occlusions between real and virtual objects. To maintain consistency in joint tracking of a user and precise coordination between the real space and virtual objects, we substituted ZED mini inside-out tracking for Vive Pro headset's tracking.

{\bf Full-body and Finger Tracking.}
We used six Vive tracking devices to capture a user's joint positions in real-time (Vive Pro headset, two trackers on each wrist, one on the front of the pelvis, and two on each foot). Each tracker records 6 DoF global position and orientation of the corresponding joint. Before running the telepresence system, users go through the calibration stage for scale adjustment with the virtual avatar model. Tracked global transformations are converted into local transformations with respect to the user's local coordinate system and transferred to the remote avatar. 

For a remote avatar's accurate finger action, which is essential for non-verbal communication and deictic gesture, a user wears Noitom Hi5 VR Gloves. The user's finger joint rotations are captured from the gloves and applied directly to the remote avatar's character rig. 

{\bf Networking.}
Communication between two distant systems is achieved by Photon Unity Network (PUN) framework. We set the desired send rate to match the system's frame rate (60fps) for minimum latency between the systems. The channel is set as a reliable connection because output placement from the placement module and resulting full-body avatar pose are heavily affected by a small difference in transferred feature values and transformations. The audio is captured from the built-in microphone in Vive Pro HMD and transferred with the PUN Voice framework.

{\bf Placement Inference Module.}
Avatar placement module is implemented with Python/Tensorflow. The module receives inputs as a feature vector from the remote system and outputs corresponding avatar placement in the local system. The system and the module communicate via Python socket connection. For real-time inference of avatar placement, the grid map of local space is divided into several partial maps for parallel processing, and user/avatar-independent features are precomputed for each grid position and direction. The average inference time is 400ms (300ms for grid-level optimization, 100ms for PSO) for each placement.

\begin{figure}[t]
    \centering
    \begin{subfigure}{0.90\columnwidth}
        \includegraphics[width=1.0\columnwidth]{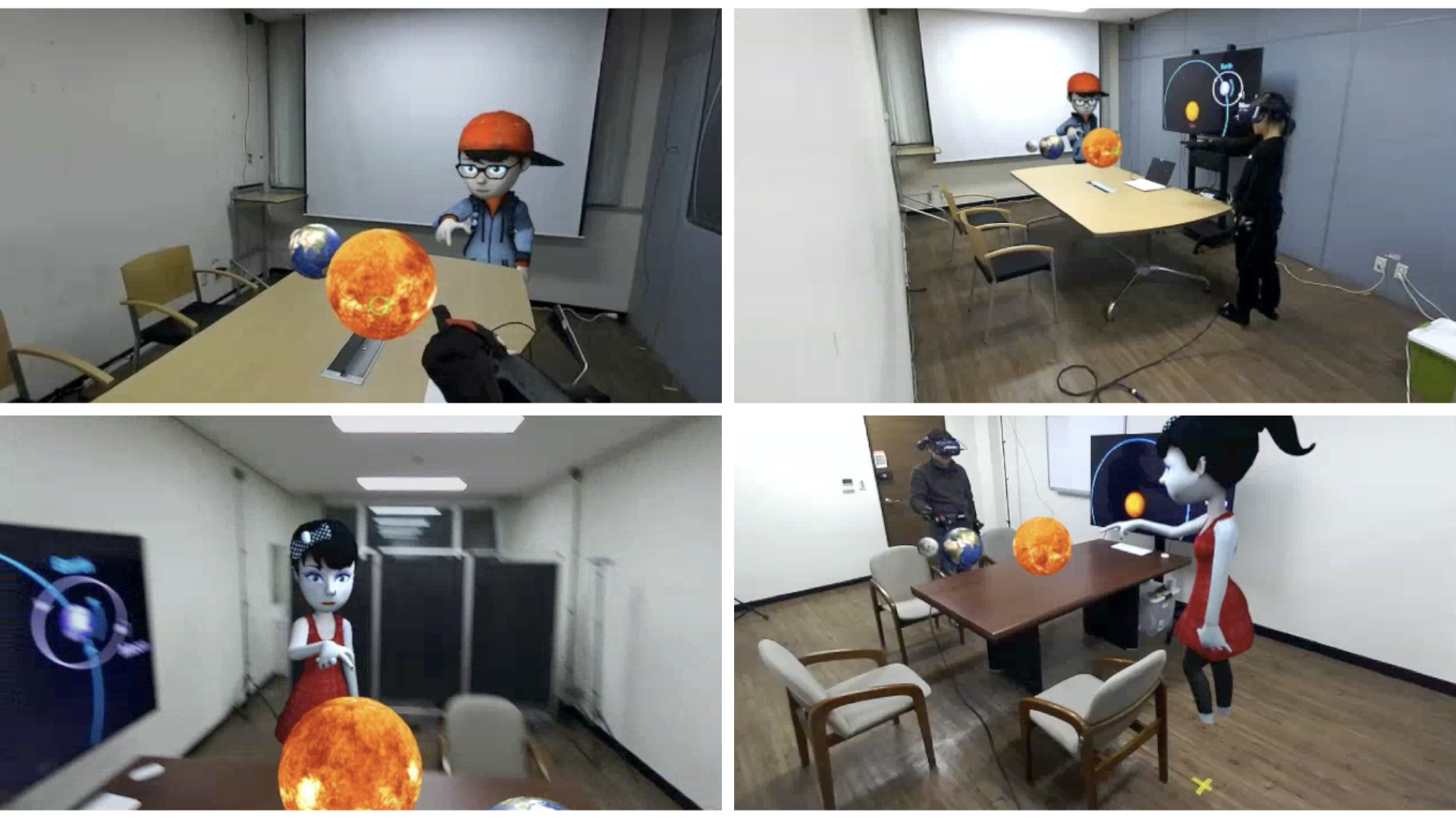}
        \caption{\textbf{Interactive learning:} Face-to-face deictic interaction in an office. }
        \label{fig:learning}
    \end{subfigure}\\
    \begin{subfigure}{0.90\columnwidth}
        \includegraphics[width=1.0\columnwidth]{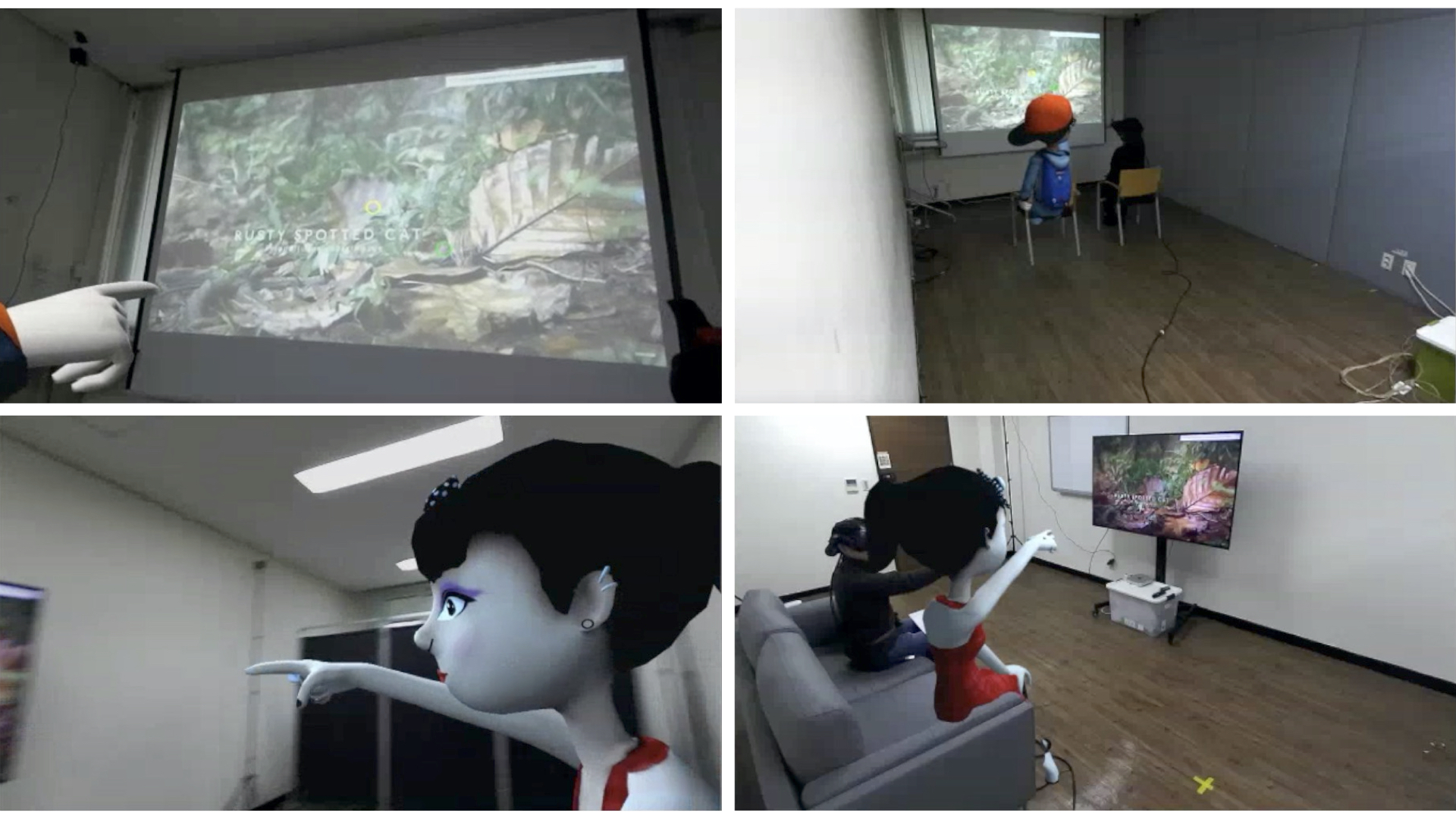}
        \caption{\textbf{Watching Video Together:} Side-by-side sitting in a living room. }
        \label{fig:watchingTV}
    \end{subfigure}    
    \begin{subfigure}{0.90\columnwidth}
        \includegraphics[width=1.0\columnwidth]{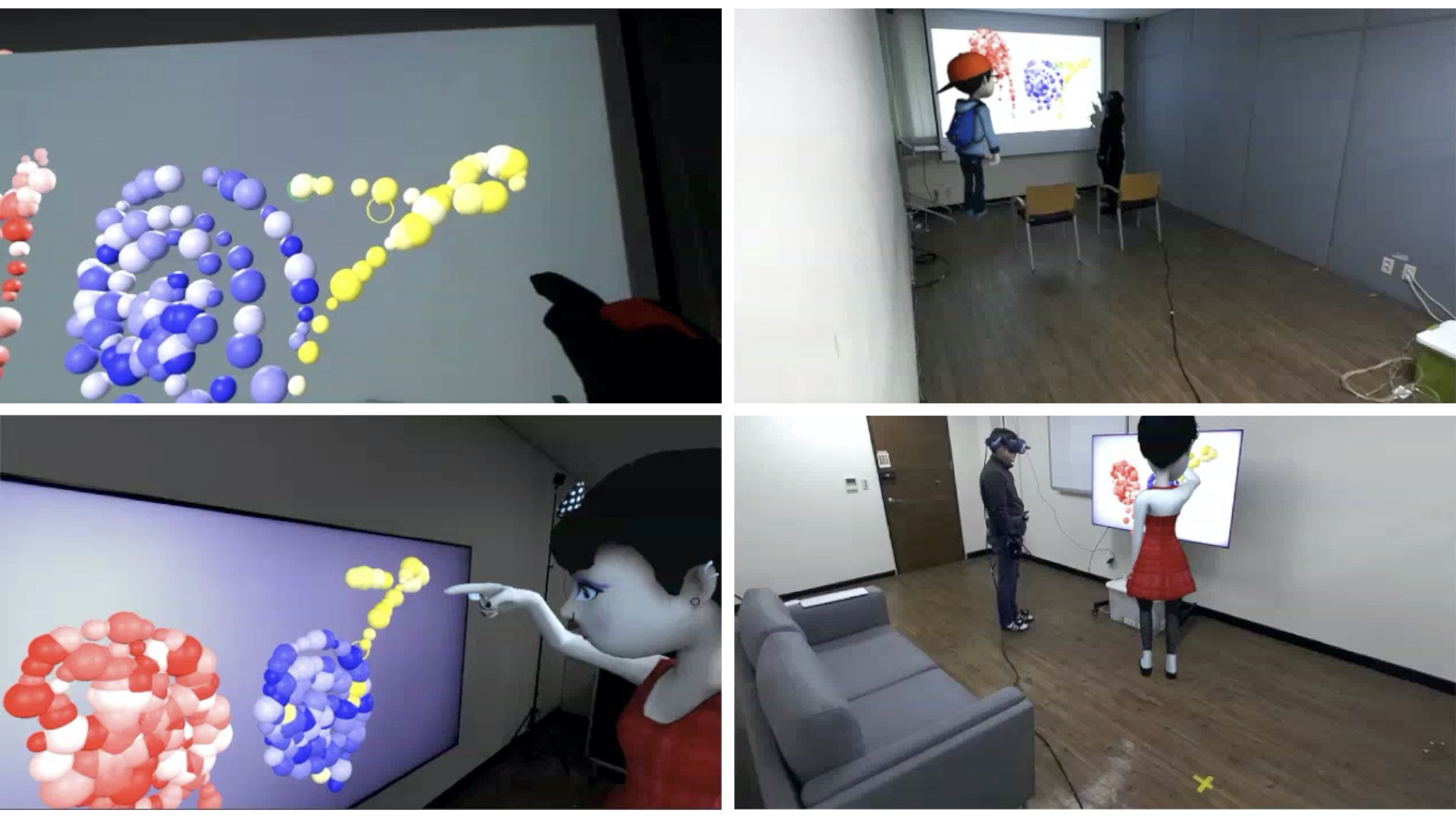}
        \caption{\textbf{Digital Painting:} Collaborative painting on a shared canvas.}
        \label{fig:painting}
    \end{subfigure}    
    \caption{Screenshots of tasks. In each figure, egocentric (left) and perspective (right) views of a space are shown in parallel.}
    \label{fig:tasks}
\end{figure}

\section{Scenario}
\label{scenario}
We expect many possible applications for our system in broad areas such as business meetings, remote education, and social gathering. Here, we suggest two representative scenarios in an office room and a living room. 

{\bf Office Room.} In a typical office room, equipped with a chair, a table, and a large screen, the users can utilize the furniture, communicate with a partner using full-body gesture and point to a presentation material on a screen during a business meeting. Moreover, a shared object can be augmented on a table for interactive learning or group design activity.

{\bf Living Room.} In a living room, furnished with a couch, a set of chairs, and a large TV, a user can sit together with a friend or a family member and watch TV while pointing on the contents and share stories with natural non-verbal reactions from full-body. Also, the uptake of the full-body avatar representation and finger tracking enables numerous interactive applications for entertainment purposes, such as collaborative painting or a classical board game using augmented objects.

\section{Exploratory User Study}
Having proposed major functionalities of the system for telepresence scenario in two remote rooms, we conducted an exploratory user study to investigate the effect of our system on user experience, the opportunities for developing novel interaction techniques, and future challenges of our systems. The example demo of tasks performed by the authors is shown in Figure \ref{fig:tasks} and the supplementary video.

\subsection{Procedure and Tasks}
We recruited 6 pairs of participants (a total of 12 participants: 7 males, 5 females: age 25 to 35) from a pool of students in a local university community.  Each pair of participants knew each other prior to the study, and all participants had experience with VR or AR interfaces before. In each study session, two participants were placed in different rooms and asked to perform interactive learning tasks in the office room setting and social interaction tasks in the living room setting by following the suggested scenarios in the previous section. We did not inform the participants about the applied retargeting techniques and counterbalanced the order of the tasks using the balanced Latin square design. The whole procedure took approximately 45 minutes, including a 10-minute warm-up session. After the session, we conducted semi-structured interviews to obtain insight from the users’ experience and find out challenges and potential usage scenarios.    

\subsubsection{Interactive learning task in office room} 
To explore how people communicate with pointing gestures on a flat-screen or a 3D object, we asked participants to converse about the Sun, Moon, and Earth displayed on the screen while sitting in the chairs, and then they were asked to stand around a table for a deeper discussion with 3D dynamic objects \cite{SolarSystemGithub} augmented above the center of a table for both rooms. Participants were encouraged to use pointing gestures during the conversation.

\subsubsection{Social interaction task in living room} 
To probe user experience during a social activity while using our system, the participants were requested to watch a short video clip \cite{BBCCat} together while sitting side by side in a chair or a couch, and to perform digital painting after standing in front of a screen. During each task, the participants were requested to talk about the contents of the video clip and their drawings. 

\subsection{Results and Discussion}
All 12 participants stated that they had a positive experience interacting with an avatar of the other party, but the discrepancies arose from the dissimilar spaces remained especially during the teleportation of the avatar from the locomotion state. We grouped major elements of gathered responses into the following 4 categories and discussed the drawbacks of the system below. 

\subsubsection{Presence and Co-presence During the Task }
Utilizing the full-body avatar in the space with real furniture helped the participants focus more on the conversation than 2D video conferencing. P3: \textit{``While using the video conferencing application, I tended to slack during the talk by others. [...] I felt like my partner was looking at me in my place. It would be effective to participate in online lecture using this system in a classroom setting.''} P7 mentioned that \textit{``I felt the strong presence of a partner and I also liked the reaction from the body gesture.''} and P8 added that \textit{``This system enabled spatial interaction with environment context matching.''} 

Participants also felt strong co-presence while looking and pointing at the augmented object and painting on the screen. P8: \textit{``My partner and I were able to point at the Earth together, and the collaborative painting was a hilarious experience.''} P9: \textit{``I can see where the avatar is looking at, which was good during the communication.''} 

Nonetheless, some participants were dissatisfied with the sudden jump of the avatar teleportation after the Walk-In-Place, which disturbed the co-presence. P10: \textit{``I was surprised by the abrupt teleportation of avatar and started to wonder what was going on in the remote space.''}

\subsubsection{Effectiveness of Gesture Retargeting}
Overall, the gesture retargeting was functional and most participants did not spot a major problem during the task. P1: \textit{``I did not notice that the gesture was being retargeted, and I was surprised that the screen size of two rooms was different after the task is done.''} P7 also mentioned that \textit{``My partner seemed to be pointing at the right context and I was not focusing on the pointing gesture all the time.''}

However, the finger tracking problem of the IMU sensor glove and interpretation error described in \cite{Mayer:2020:IHA} were noticeable during the task. P12: \textit{``Self-pointing gesture was a bit off from the circle-cue on the screen and the movement of the avatar's fingers was not natural sometimes.''} 

Another drawback is that if the distance between two interaction targets in the local space is too close, whereas the distance between the corresponding targets in a remote space is too far, the retargeted deictic gesture would be largely different from the original motion of the user. The avatar placement algorithm does not consider the relationship between the interaction targets, and it can be further analyzed in future work.

\subsubsection{Curiosity about the Local Space of Avatar}
Many participants maintained a focus on the avatar of the remote person and were not aware of the remote person's real space, especially during the interaction. P2: \textit{``I wasn't curious about the remote space at all. I remained strong focus on the screen or the avatar.''} P6: \textit{``I focused only on the avatar, which seemed to be being at my place and wasn't curious about the remote space during the interaction.''}

However, some participants noticed the discrepancy between two spaces during the task due to the avatar's sudden jump or the absence of spatial sound support. P4: \textit{``After a few avatar placements, I started to be worried that different configurations between two rooms might affect interaction. It would be nice to have a glimpse of the remote room.} P5: \textit{``Due to the absence of spatial sound, I was confused about the actual location of partner and felt like doing a voice chatting.''}

\subsubsection{Possible Extensions}
Many participants suggested practical applications to our system during the interview. P4 mentioned that \textit{``This system would be ideal for making a presentation material as a group. I can point on the screen and look at my partner to observe a body reaction as well as voice chat while designing a slide together in remote places.''} Our system is also suitable for social application as described by P7 that \textit{``I would like to do blind dating with this system. I can meet many different people to find a match in a short period of time.''} P2 envisioned a novel way of online shopping by commenting that ``I would like to shop clothes with my sister in home town using this system. Pointing with a full-body expression on a cloth represented as a realistic 3D object as well as voice chat will definitely expedite our buying decision.''

\section{Conclusions and Future Work}
We have presented a novel full body avatar-based telepresence system for dissimilar spaces. The proposed system enables the users in dissimilar spaces to interact with a remote user and a shared virtual object as well as existing objects in their rooms. The result of user study showed that our system was functional for remote communication and collaboration. Moreover, we found several possible ways to improve our system, based on feedback from the participants.

First, the locomotion state can be further developed for the interaction between users. WIP method does not relay the user's rich information in remote space in terms of the path or route in between placements. Moreover, a sudden jump of an avatar by the teleportation disturbed the participants' co-presence in remote space, especially when the avatar's final destination is closer to the personal space around the user. One possible approach is to retarget walking motion from the initial location to the final destination to slowly build up the spatial context of avatar relocation to the user. This is challenging future work that deals with a user's attention prediction and a time-distance discrepancy between two different paths.

Second, the locomotion state can be further divided into a quasi-static movement state and a concurrent movement state. During the solo state, a user occasionally takes a step toward an object or a partner. Also, two users can move to different locations simultaneously depending on the conversation context or direct each other's position manually. A user can try to direct the remote user to move, solely based on one's own sense of bearing in local space, inevitably disregarding the unknown spatial configuration of the remote space. Careful consideration of these states can further improve user experience in terms of co-presence and immersion.  

Third, we should also invest in animating more natural avatar transitional motion for gesture retargeting. Confusing non-verbal cues are evident to the user, especially when the avatar state changes from interaction back to solo due to a mismatch in the original relationship of avatar placement and communication context. We plan to suggest and evaluate additional measures to retarget a wider range of user gestures. Users sometimes expressed difficulty when observing and interpreting the avatar's deictic gestures. Such feedbacks were common in side-to-side configurations where users had more difficulty noticing non-verbal cues of the avatars. In pointing scenarios where the correct interpretation of avatar pointing directly impacted communication of information, users reported that occlusion of avatar motion, the position of a pointed object with respect to avatar placement led to ambiguities in interpreting deictic gestures. We believe augmented awareness cues \cite{cues} such as pointing cursors or pointing lines can help us understand the other user's point of focus, and further study on utilizing them without cluttering the MR environment can be a valid research topic.

Participants also suggested some possible improvements that were not directly related to the current research focus but still worth mentioning. Several participants reported that an absence of spatial sound implementation made them feel they were participating in a non-stereo voice chat, which disturbed the feeling of co-presence on some occasions. Also, the use of an avatar that resembles the user was suggested to enhance the user experience.   

With the steady advancement of these future research topics, we will be able to take a step closer to a truly immersive co-presence experience that overcomes the distance between people.

\acknowledgments{
This work was partly supported by Samsung Science and Technology Foundation (SRFC-IT1701-14).}

\bibliographystyle{abbrv-doi.bst}

\bibliography{template}
\end{document}